\documentclass[11pt]{article}
\usepackage{amsmath, amsfonts, amssymb}
\usepackage{graphics}

\begin{document}
\title{On the Nature of Dark Matter and Dark Energy}
\author{Yu.A.Baurov$^1$, I.F.Malov$^2$\\
1) Central Research Institute of Machine Building,\\ 141070,
Pionerskaya, 4, Korolyov, Moscow Region, Russia;\\
2) P.N.Lebedev Physical Institute, Russian Academy of Sciences,\\
117924, Leninski pr., 53, Moscow, Russia. }

\date{}
\maketitle
\begin{abstract}
It is shown that some problems connected with dark matter and dark
energy can be solved in the framework of the byuon theory.
\end{abstract}
\section{Introduction}

Some hard problems have appeared in astrophysics during the last
dozens of years. Observations show that approximately $3-4 \%$ of
the cosmological energy density is accounted for by baryons,
$25-27 \%$ by "the dark matter" and the reminder by "the dark
energy" (see, for example, [1-2]).

There are some evidences for the existence of dark matter (DM) and dark energy (DE).

Here we enumerate basic ones only.

1) In 1937 F.Zwicky measured velocities of galaxies in the Coma cluster and concluded that the total mass of this cluster must be much more than observable one to prevent the escaping of investigated galaxies from the cluster.

2) The summarized mass of the observed gas and galaxies in the number of clusters is not enough to keep them inside of the cluster.

3) The gravitational lensing  by clusters of galaxies gives the mass of such lens much more than observable one.

4) The rotation curves of galaxies [3] show that the total mass of the individual galaxy is approximately one order higher than the mass of gas and all stars observed in this galaxy.

5) The observations of supernovae in distant galaxies (see, for example, [4]) show that our Universe expanses with an acceleration, and there is a source causing such type of expansion.

The nature of dark matter and dark energy  is unknown up to now.

\section{Dark matter}

The DM is not observed as shining matter and must have very weak electromagnetic interactions. It must be approximately collisionless and non-relativistic.

The DM is not primarily baryonic. The calculated amount of deuterium should be much smaller than observed one if the average baryon density was an order of magnitude higher than the modern value ($\sim 0.3$ per cubic meter).

The mass interval for the possible candidates in DM is huge (from $10^{-22}$ eV to $10^6 M_{\odot} \approx 10^{72}$ eV).

Let us discuss the most probable candidates in DM.

1) Axions, light pseudo-scalar bosons [5-6] with mass  $\mu e V \lesssim m \lesssim m e V$. They could be detected by resonant axion-photon conversion in a magnetic field  [7-8].

2)  Neutrinos. Some laboratory experiments and cosmological restrictions give the mass interval for all kinds of neutrinos

$$
50 m e V \lesssim \sum m_{\nu} \lesssim 0.7 e V, or
$$

$$
0.0005 < \Omega_{\nu} h^2 < 0.0076,
$$

where $\Omega_{\nu} = \rho_{\nu} / \rho_c$, $h = H_0 / 100 km / sec / M p c$, $\rho_c  = 3 H_0^2 / (8 \pi G)$ is the critical density of the Universe, $H_0$ is the Hubble constant.

Super-symmetric theories put bosons and fermions in common multipletes. They give some possible candidates in DM.

3) The super-partner of the graviton, gravitino with the spin $3/2$ [9].

4) Neutralinos. These are the four spin $1/2$  Majorana fermion super-partners of the neutral gauge and Higgs bosons ($\chi^0_{1-4}$) [10]. There are also two charged  Dirac fermion super-partners of charge gauge and Higgs bosons, charginos ($\chi^{\pm}_{1-2}$).

5) Axinos, a spin $1/2$ partner of the axion  [11].

6) Non-topological solitons, Q-balls [12].

7) If our four-dimensional space-time is embedded in a higher dimensional space, the Kaluza -- Klein excitations of Standard Model states along the orthogonal dimensions may be as DM candidates [13].

8) Objects of many dimensions (branes) are described in string theories. Their fluctuations have been considered as particles (branons) which could be DM candidates [14].

9) DM could be an ordinary matter in the mirror world where the only communication is gravitational. In this case our Universe and a mirror universe are two branes in a higher dimensional space [15].

10) At the last stages of inflation gravitational interactions can produce a lot of weakly interacting massive particles which for mass scales of $10^{13}$ GeV could account for DM [16].

11) Primordial black holes have been considered as candidates in DM as well [17].

So, as observations give, baryons provide approximately $4 \%$ of
DM, neutrinos $\sim 0.3 - 3 \%$ of it. The rest ($20-25 \%$) is a
non-baryonic DM. The nature of this part of DM is unclear. There
are many problems with theoretical foundations and experimental
evidences of the existence of particles mentioned above and
described in cited papers.

\section{Dark energy}

   The nature of dark energy is much more unclear than that of DM. It is necessary for it to have the equation of state of the following form (see, for example, [18]):

\begin{equation}
P = w \rho
\end{equation}

where $p$ is pressure and $\rho$ is the energy density. The most probable value of the parameter  $w$ is approximately $-1$, as follows from the known observations. This implies that the energy density of such substance  is constant and corresponds to the flat universe, i.e. the curvature K of the spatial sections (slices at constant cosmic time) is equal to zero. DE causes the acceleration of the expansion of our Universe. Fig.1 shows the sum (solid line) of two potentials: The usual (negative) gravitational potential $\varphi_1$ (broken line) causing the attraction of two bodies and  positive constant potential $\varphi_2$  giving the repulsion at large distances ($r > r^*$).

One of the possible sources of DE is "quintessence" [19], a scalar field $\Phi$ rolling slowly in a potential. Most quintessence models give for such scalar fields

\begin{equation}
m_{\Phi} c^2 \sim 10^{-33} \, eV
\end{equation}

In quantum field theory light scalar fields are hard to understand. In any case these fields give rise to long-range forces which should be observable, and it is surprisingly  why such quintessence field has not been detected up to now.

There are many problems with other models of DE (see, for example, [18]).

In this paper we shall try to explain DM and DE in the framework of the byuon theory. First of all we will describe briefly the foundations of this theory.

\section{Basic axioms and hypotheses. Space, time, and physical vacuum in the light of the byuon theory}

Any theory begins with axioms, that is, with basic postulates accepted without proofs.

Thus, let us assume that there are no space, no time, no world of
elementary particles from which all physical bodies consist, but
there is a mathematical object, a byuon $\mathbf{\Phi}(i)$
[20-25], being unobservable in itself and characterized by
discrete states (i.e. numbered by the series of natural numbers)
having inherent "vectorial" property. The expression for
$\mathbf{\Phi} (i)$ is
\begin{equation}
\mathbf{\Phi} (i) =
\begin{cases}
[\mathbf{A}_g x (i)],\\
- \sqrt{-1} [\mathbf{A}_g x (i)]\\
\end{cases}
\end{equation}

where $x (i)$ is "length" of the byuon, a real (positive or
negative) value depending on the index $i = 0,1,2, \ldots k,
\ldots$ Index $i$ is a quantum number for $\mathbf{\Phi} (i)$
\footnote{It should be explained that the vector $\mathbf{\Phi}
(i)$ is not an ordinary vector in some space but an object with
"inner" vectorial properties that are manifesting themselves when
the value $x (i)$ changes in the process of physical space
formation.}.

The quantity $A_g$ is some internal potential being equal in magnitude to the cosmological vectorial potential $\mathbf{A}_g$, a new fundamental vectorial constant introduced in Refs. [26-27] ($A_g \approx 1.95 \cdot 10^{11} \, Gs \cdot cm$).

Thus, $\mathbf{\Phi} (i)$ can take both real and pure imaginary
values.

The whole set $\Phi (i)$ forms a one-dimensional space $R_1$ in index $i$.

According to this conception, by the discrete time is meant, for
the byuon, a discrete change in the index $i$ (its increase or
decrease). In connection with the discrete time, a quantum of time
$\tau_0$ and quantum of space  are introduced in the
one-dimensional discrete space $R_1$ formed by byuons ($\tau_0
\approx 0.9 \cdot 10^{-43} \, s, \tilde x_0 \approx 2.8 \cdot
10^{-33} \,cm$). The distance between byuons is defined therewith
as a difference in their lengths $x(i)$. The space $R_1$ is
discrete by definition.

Since the space $R_1$ is discrete, one of methods of parametrization of $x(i)$ is $x(i) =\tilde x_0 \cdot i$, or $x(i) = -\tilde x_0 \cdot i$.

\textbf{Statics}. In the set $\{ \Phi (i) \}$, there are meant no
static states with time $t > \tau_0$.

\textbf{Kinematics}. Depending on whether the length $X(i)$
positive or negative, decreases or increases in magnitude, free
byuons (i.e. not interacting one with another) can be only in one
of the four so called vacuum states $(VS) \, II^+,ÿI^+,ÿI^-,
II^-$.

Introduce the following definitions.

1. A free byuon is in the state $II^+$ if its positive length
discretely, in a quantum of time $\tau_0$, increases by a quantum
of distance $\tilde x_0$ with the speed of propagation (increase
in length)   $c = \frac{\tilde x_0 - 0}{\tau_0} = c_0$ ($c_0$ is
the light speed).

2. A free byuon is in the state $I^+$ if its positive length
discretely, in a quantum of time $\tau_0$, decreases by $\tilde
x_0 $. In this case $c = \frac{0 - \tilde x_0}{\tau_0} = -c_0$.

3. A free byuon is in $II^-$ if the modulus of its negative length
increases by $\tilde x_0 $ in time $\tau_0$ with $c =
\frac{-\tilde x_0 - 0}{\tau_0} = -c_0$.

4. A free byuon is in $I^-$ if the modulus of its negative length
discretely, in time $\tau_0$, decreases by $\tilde x_0 $. In this
case $c = \frac{0 - (-\tilde x_0)}{\tau_0} = c_0$.

From the definition of byuons it is seen that they are in
perpetual dynamics of generation and annihilation, extension and
contraction. The collection of free (not interacting) byuons in
$VSs \, II^+, \, I^+, \, I^-, \, II^-$ forms physical vacuum of
the one-dimensional space $R_1$ of index $i$ (about properties of
$R_1$ will be said below). Recall however that in this model of
physical vacuum, time is a sequence of events of byuon generation
(extension) and "collapse" (contraction). There corresponds to
each byuon its own count of time measured by the natural number
series. One of the two directions of the one-dimensional space
$R_1$, coincident with that of a byuon with the maximum $x(i)$ in
$VS \, II^+$, is taken for the positive direction of the vector
$\mathbf{\\A_g} $  and $\mathbf{\Phi} (i)$.

The average magnitudes for  byuons being in the above described
$VSs$ at maximum $i=k$, are determined from the following
expressions (see (3), and [20-22]):
\begin{equation}
\begin{array}{rcl}
\bigl[A \cdot X \bigr]_{II^-}^{i+I}  = {A_g} \left[ \frac{2i + 1}{2} \cdot \frac{const_1}{k \cdot A_G} \right] & X > 0, \\
\bigl[A \cdot X \bigr]_{I^+}^{i} =  -\sqrt{-1} \, \cdot {A_g} \left[ \frac{2i + 1}{2} \cdot \frac{const_1}{k \cdot A_G} \right] & X > 0,\\
\bigl[A \cdot X \bigr]_{II^-}^{k-i}  =  \sqrt{-1} \, \cdot {A_g} \left[ \frac{2(k - i) - 1}{2} \cdot \frac{const_1}{k \cdot A_G} \right] & X < 0,\\
\bigl[A \cdot X \bigr]_{I^-}^{k-i-1}  =  -{A_g} \left[ \frac{2(k - i) - 1}{2} \cdot \frac{const_1}{k \cdot A_G} \right] & X < 0,\\
\end{array}
\end{equation}

where $const_1 = \frac{\sqrt{h c_0}}{4 \sqrt{3}} \cdot \frac{h c_0}{e_0^2}$ is some constant. As we will see later $h$ is equal to the Plank constant and $e_0$ is the electron charge.

Assume that for the byuons with the length greater than $\tilde x_0$, only contact interactions are realized, by which we will mean existence of at least  two byuons at a quantum of space $R_1$.

\textbf{Hypothesis 1}. Assume the observable three-dimensional
space $R_3$ to appear as a result of minimization of the potential
energy (PE) of byuon interaction in the one-dimensional space
$R_1$ formed by them. We construct PE from the taking into
consideration of dimensions. More precisely, the space $R_3$ is
fixed by us as the result of this byuon dynamics. In the space
$R_3$ therewith the dynamic processes for objects with the
residual positive potential energy of byuon interactions
originate, and in consequence, the wave properties of elementary
particles arise.

The proposed hypothesis requires to develop a mathematical model
basing on a new algebra of probabilistic events since the
elementary events (a discrete decrease ($\overline{\\D}$) or
increase (D) in the length of byuon) are assumed to be
probabilistic in character. Hence for the byuons of the minimum
length we may say about the existence, with certain probability,
of the events ($\overline{\\D} \cdot D \ne 0$. Note that in Ref.
[9] an algebra of events is given, being a development of the
Boolean algebra with the proviso that $\overline{\\D} \cdot D =
1$. For the deterministic approach used in [28], the event
$\overline{\\D} \cdot D \ne 0$ is illogical by von Neumann, but in
the probabilistic space of events the existence of $\overline{\\D}
\cdot D \ne 0$ is possible.

In this paper only the physical statement of the problem will be considered, and results of evaluations made in support of the hypothesis advanced, will be given.

The space $R_1$ is formed from the set of byuons in such a manner that at its $i$-th point there exist all the byuons with the lengths smaller than $X(i)$ or equal to $X(i)$ for $X(i) > 0$, and those with the absolute values smaller than $X(k-i)$ for $X(k-i) < 0$, where $k$ is some period in $i$.

The assumption that two neighboring byuons (the $i$-th and
$(i+1)$-th; $(i+1)$th and $(i+2)$th etc.) being in vacuum states
$II^+$ will interact, is unreasonable since in this case the
definition of byuons for this $VS$ would be violated at the point
of interaction. Such interaction is possible only between the
$i$-th and $(i-k)$th byuons in the state $II^+$ if they form a
"loop" in the space $R_1$ (by the "loop", the periodicity of the
process in $i$ is implied), i.e. the two byuons $II_i^+$  and
$II_{i-k}^+$   will be observed simultaneously at one point of the
space $R_1$. The least possible value of $k$ is $k=3$. In Fig.2
shown is the interaction of byuons in the vacuum states $II_1^+$
and $II_4^+$   (the smallest loop). The byuons in the state $II^-$
interact likely.

The byuon states $I^+$ and $I^-$ can occur only if the byuons have
already been in $VS \, II^+$ and $II^-$, respectively. The vacuum
states $I^+$ and $I^-$ as if "occupy" $VSs \, II^+$ and $II^-$. At
maximum positive potential energy of byuon interaction there
exists a single variant of "occupancy" (Fig.ÿ3).

The probability of the minimum four-contact interaction of the
neighboring in i byuons $I^+ II^+ \timesÿI^- II^-$ ("$\times$"
symbolizes interaction) with randomly appearing states $I^+$ and
$I^-$, is equal to 1/16 [20-22]. That is quite understandable when
analyzing possible four-contact interactions (see Fig.4)). All
other possible variants of the four-contact interaction are
unobservable either because one cannot introduce them without
violating the definition of byuons or in view of imaginary energy
of such interaction.

Note once more that there exist only two directions in the
one-dimensional world, the first of which corresponds to
increasing $i$ index for byuons with $X > 0$ (vacuum state
$II^+$), and the second corresponds to decrease in $i$ for such
byuons (vacuum state $I^+$). These directions are coincident with
those for byuons with $X < 0$: $II^-$ with $I^+$, and $I^-$ with
$II^+$. It is clear from above definitions that the byuon with the
maximum length $X(i)$ in $VS \, II^+$ determines the positive
direction, and directions of other byuons are correspondent with
it.

The four-contact interaction of byuons is realized within a time
$\tau = \tilde \tau_0$ only at points $D$ of the $R_1$-space
(Fig.3), i.e. at the points where introducing an interaction with
$PE > 0$ is possible. In Fig.3 the arrows corresponding to byuons
show directions of decrease or increase in their lengths relative
to the origin of the coordinates introduced, for example, where $i
\rightarrow  0$ (in its direction the absolute value of the byuon
length is decreased (states $I^+$ and $I^-$), and it increases in
the opposite directions (states $II^+$ and $II^-$)). At the points
$A$ in Fig.3, the coordinate denoting place (time) of byuon
interaction cannot be fixed because of violating, in such a case,
the definitions of the byuon states (in one quantum of the
$R_1$-space within a time $\tau_0$, the byuons with $II^+ I^+ I^-
II^-$ should not be present). It is assumed that before the origin
of $VS \, II^+$ with the minimum length ($i=1$), the byuon vacuum
states $II^-$ and $I^-$ with any possible lengths are already in
existence.

Propagation of byuons in $VSs \, II^+ I^+$ and $I^- II^-$, the interaction between which occurs with imaginary energy (see below ), presents two wave-like processes (see below) directed towards each other at $X(i) > 0$ and $X(i) < 0$, respectively. These processes are unobservable. A really observable signal can be transmitted by means of such processes only in the four-contact byuon interaction $II^+ I^+ \times I^- II^-$.

Obtain an equation characterizing propagation of the four-contact interaction of byuons in $R_1$. Introduce functions of index $i$, characterizing the origin of such or another $VS$ by byuons: $\Psi_{II^+}^{i+2}, \Psi_{II^-}^{k-i}$, determining the processes of byuon length magnitude origin and increase at positive and negative $X(i)$, respectively; $\Psi_{I^+}^{i}, \Psi_{I^-}^{k-i-2}$, determining the processes of byuon length magnitude cancellation and decrease at positive and negative $X(i)$, respectively.

The physical sense of the introduced functions consists in that their product determines the probability of two-contact interaction of byuons (for example, $\Psi_{II^+}^{i+2} \cdot \Psi_{II^+}^{i+2-k}$ determines the probability of interaction of byuons $[A \cdot X]_{II^+}^{i+2}$ and $[A \cdot X]_{II^+}^{i+2-k}$), the product of four functions determines the probability of four-contact interaction, the product of eight functions gives the probability of eight-contact interaction. These products should be positive, and in this case only they can describe an observed event.

The probability of a single event is no greater than 1.

Depending on which range is $i$ in ($0 \le i < k, \, k < i < N k,
\, N k < i < N k P$ where $k,  N, P$ are the assumed periods in
$i$) various types of contact interactions between byuons may be
introduced, and hence the normalization of the introduced
functions should be dependent on $i$.

Let us normalize the introduced functions for the case $0 \le i < k$ in the following manner

\begin{subequations}

\begin{equation}
\sum_{\xi=0}^{(NkP-k)/2} \sum_{j=0}^{j=1} \Psi_{II^+}^{j+2} \cdot
\Psi_{I^+}^{j} \cdot \Psi_{II^-}^{NkP-j-2-2\xi} \cdot
\Psi_{I^-}^{NkP-j-2-2\xi} = \frac{NP}{2},
\end{equation}

\begin{equation}
\sum_{\xi=1}^{NP-1} \sum_{j=0}^{j=1} \Psi_{II^-}^{NkP-j} \cdot \Psi_{II^-}^{NkP-j-2-\xi k} = P ,
\end{equation}

\begin{equation}
\sum_{\xi=0}^{(NkP-k)/2} \sum_{j=0}^{j=1} \Psi_{II^+}^{j+2} \cdot \Psi_{I^-}^{NkP-j-2-2\xi} = \frac{NP}{2} ,
\end{equation}

\begin{equation}
\sum_{\xi=0}^{(NkP-k)/2} \sum_{j=0}^{j=1} \Psi_{I^+}^{j} \cdot
\Psi_{II^-}^{NkP-j-2-2\xi} = \frac{NP}{2}
\end{equation}

\end{subequations}

When normalizing, it was taken into account that within a period
in $i = k$, one four-contact interaction occurs with probability
1.

Obtain an equation in terms of $\Psi$-functions, describing the propagation of four-contact interaction of byuons. For that we may write the following relationships as to the origin of $VSs \, II^+ (f_{II^+}), I^+ (f_{I^+}), II^- (f_{II^-}), I^- (f_{I^-})$ depending on certain $VSs$ of the byuons neighbouring in the index $i$:
\begin{equation}
\begin{array}{rcl}
\Psi_{II^+}^{i+2} & = & f_{II^+} [\Psi_{II^+}^{i+1}, \Psi_{I^+}^{i+2}, \Psi_{II^-}^{NkP-i}, \Psi_{I^-}^{NkP-i-2}],\\
\Psi_{I^+}^{i} & = & f_{I^+} [\Psi_{I^+}^{i+1}, \Psi_{II^+}^{i}, \Psi_{II^-}^{NkP-i}, \Psi_{I^-}^{NkP-i-2}],\\
\Psi_{II^-}^{NkP-i} & = & f_{II^-} [\Psi_{II^-}^{NkP-i-1}, \Psi_{I^-}^{NkP-i-1}, \Psi_{II^+}^{i+2}, \Psi_{I^+}^{i}],\\
\Psi_{I^-}^{NkP-i-2} & = & f_{I^-} [\Psi_{I^-}^{NkP-i-1}, \Psi_{II^-}^{NkP-i-2}, \Psi_{II^+}^{i+2}, \Psi_{I^+}^{i}].\\
\end{array}
\end{equation}

Assuming only linear dependences in expressions (6) as well as equiprobability of $VSs$ of byuons neighbouring in $i$, we obtain the following equations for $\Psi$-functions of four-contact interactions of byuons:
\begin{equation}
\begin{array}{rcl}
\Psi_{II^+}^{i+2} & = & \Psi_{II^+}^{i+1} - \Psi_{I^+}^{i+2} - \Psi_{II^-}^{NkP-i} + \Psi_{I^-}^{NkP-i-2},\\
\Psi_{I^+}^{i} & = & \Psi_{I^+}^{i+1} - \Psi_{II^+}^{i} + \Psi_{II^-}^{NkP-i} - \Psi_{I^-}^{NkP-i-2},\\
\Psi_{II^-}^{NkP-i} & = & \Psi_{II^-}^{NkP-i-1} - \Psi_{I^-}^{NkP-i-1} - \Psi_{II^+}^{i+2} + \Psi_{I^+}^{i},\\
\Psi_{I^-}^{NkP-i-2} & = & \Psi_{I^-}^{NkP-i-1} - \Psi_{II^-}^{NkP-i-2} + \Psi_{II^+}^{i+2} - \Psi_{I^+}^{i}].\\
\end{array}
\end{equation}

From the first and second pairs of Eqs.(7) it is easy to obtain, respectively, the following equations:

\begin{equation}
\Delta [\Psi_{II^+}^{i+1} +\Psi_{I^+}^{i+1}] + \Psi_{II^+}^{i+1} +\Psi_{I^+}^{i+1} = 0,
\end{equation}

\begin{equation}
\Delta [\Psi_{II^-}^{NkP-i-1} + \Psi_{I^-}^{NkP-i-1}] +
\Psi_{II^-}^{NkP-i-1} + \Psi_{I^-}^{NkP-i-1} = 0,
\end{equation}

where $\Delta$ denote the second finite differences in index $i$.

It is seen from Eqs.(8) and (9) that the process in $i$ is of
oscillatory character for the functions $([\Psi_{II^+}^{i+1}
+\Psi_{I^+}^{i+1})$ and $(\Psi_{II^-}^{NkP-i-1} +
\Psi_{I^-}^{NkP-i-1})$. These functions determine the $A$-type
points in the space of index $i$ shown in Fig.3, i.e. the points
at which we cannot introduce interaction of byuons. The $A$-type
points in the space $R_1$ determine as if "ruptures" in $i$,
between which there exist objects with energy $E > 0$.

For the case $i \le k$, write an equation for an increment in
potential energy of byuon interaction $\Delta E(i)$, corresponding
to the occurrence of $VSs$ $II_{+}^{i+2}$ and $I_{+}^{i}$.
Minimization of $E(i)$ is assumed to be going at each step in $i$.

\begin{equation}
\begin{array}{rcl}
\Delta E (i) & = & \Psi_{II^{-}}^{NkP-i-k} \Psi_{II^{-}}^{NkP-i} E_{II^{-} II^{-}}^{NkP-i-k, NkP-i} \cos_{II^{-} II^{-}} +\\
& + & \Psi_{II^{-}}^{NkP-i-2k} \Psi_{II^{-}}^{NkP-i} E_{II^{-} II^{-}}^{NkP-i-2k, NkP-i} \cos_{II^{-} II^{-}} +\\
& + & \Psi_{II^-}^{NkP-i-3k} \Psi_{II^-}^{NkP-i} E_{II^- II^-}^{NkP-i-3k, NkP-i} \cos_{II^- II^-} + \ldots\\
& + & \Psi_{II^+}^{i+2} \Psi_{I^-}^{NkP-i-2} E_{II^+ I^-}^{i+2, NkP-i-2} \cos_{II^+ I^-} +\\
& + & \Psi_{II^+}^{i+2} \Psi_{I^-}^{NkP-i-2-2} E_{II^+ I^-}^{i+2,
NkP-i-2-2} \cos_{II^+ I^-} +\\
& + & \Psi_{II^+}^{i+2} \Psi_{I^-}^{NkP-i-2-2 \times 2} E_{II^+ I^-}^{i+2, NkP-i-2-2 \times 2} \cos_{II^+ I^-} +\\
& + & \Psi_{II^+}^{i+2} \Psi_{I^-}^{NkP-i-2-2 \times 3} E_{II^+
I^-}^{i+2, NkP-i-2-2 \times 3} \cos_{II^+ I^-} + \ldots\\
& + & \Psi_{II^-}^{NkP-i} \Psi_{I^+}^{i} E_{II^- I^+}^{NkP-i, i} \cos_{II^- I^+} + \\
& + & \Psi_{II^-}^{NkP-i-2} \Psi_{I^+}^{i} E_{II^- I^+}^{NkP-i-2,
i} \cos_{II^- I^+} +\\
& + & \Psi_{II^-}^{NkP-i-2 \times 2} \Psi_{I^+}^{i} E_{II^- I^+}^{NkP-i-2 \times 2, i} \cos_{II^- I^+} +\\
& + & \Psi_{II^-}^{NkP-i-2 \times 3} \Psi_{I^+}^{i} E_{II^- I^+}^{NkP-i-2 \times 3, i} \cos_{II^- I^+} + \ldots \\
& + & \Psi_{I^+}^{i} \Psi_{II^+}^{i+2} \Psi_{I^-}^{NkP-i-2} \Psi_{II^-}^{NkP-i} \times\\
& \times & \sqrt{E_{I^+ II^+}^{i,i+2} \cos_{I^+ II^+} E_{I^- II^-}^{NkP-i-2, NkP-i} \cos_{I^- II^-}} + \\
& + & \Psi_{I^+}^{i} \Psi_{II^+}^{i+2} \Psi_{I^-}^{NkP-i-2-2} \Psi_{II^-}^{NkP-i-2} \times\\
& \times & \sqrt{E_{I^+ II^+}^{i,i+2} \cos_{I^+ II^+} E_{I^- II^-}^{NkP-i-2-2, NkP-i-2} \cos_{I^- II^-}} + \\
& + & \Psi_{I^+}^{i} \Psi_{II^+}^{i+2} \Psi_{I^-}^{NkP-i-2-2 \times 2} \Psi_{II^-}^{NkP-i-2 \times 2} \times\\
& \times & \sqrt{E_{I^+ II^+}^{i,i+2} \cos_{I^+ II^+} E_{I^- II^-}^{NkP-i-2-2 \times 2, NkP-i-2 \times 2} \cos_{I^- II^-}} + \\
& + & \Psi_{I^+}^{i} \Psi_{II^+}^{i+2} \Psi_{I^-}^{NkP-i-2-2 \times 3} \Psi_{II^-}^{NkP-i-2 \times 3} \times\\
& \times & \sqrt{E_{I^+ II^+}^{i,i+2} \cos_{I^+ II^+} E_{I^- II^-}^{NkP-i-2-2 \times 3, NkP-i-2 \times 3} \cos_{I^- II^-}} + \ldots\\
\end{array}
\end{equation}

where $ E_{II^- II^-}^{NkP-i-k, NkP-i}, \ldots E_{I^+
II^+}^{i,i+2} $, etc. are maximum values  of potential energy of
interaction of byuons with the lengths $X (N k P $-$ i - k) < 0$
and $X (N k P - i) < 0$ in $VS \,II^-$, as well as byuons with the
lengths $X(i)$ and $X(i+2)$ in $VSs \, I^+$ and $II^+$,
respectively $\cos_{I^+ II^+} $, $\cos_{I^- II^-}$ etc. are
functions minimizing the potential energy of interaction of byuons
entering into the expressions for $E_{I^+ II^+}^{i,i+2} ... E_{I^-
II^-}^{NkP-i-2, NkP-i}$ etc. These functions are "responsible" for
the appearance of a minimum plane object and introduction of the
concept of spin (see below).

The difference in the average values of byuon lengths calculated basing upon the definition of byuons with the use of the rule of circular arrow (see Fig.5), is taken as a distance between the interacting byuons to find $\Delta E(i) > 0$. The meaning of this rule is that the said distance is calculated as
the difference in the length values of the subsequent and preceding byuons in the direction pointed by an arrow.

The lengths of byuons are:
$$
\begin{array}{rcl}
X_{II_{cp}^+}^{i+2} = \frac{2i + 1}{2} \cdot \frac{const_1}{k \cdot A_G};  & \quad &
X_{II_{cp}^-}^{k-i} = -\frac{2(k-i) - 1}{2} \cdot \frac{const_1}{k \cdot A_G};\\
X_{I_{cp}^+}^{i} = \frac{2i + 1}{2} \cdot \frac{const_1}{k \cdot A_G};  & \quad &
X_{I_{cp}^-}^{k-i-1} = -\frac{2(k-i) - 1}{2} \cdot \frac{const_1}{k \cdot A_G};\\
\end{array}
\eqno etc.
$$

The distance between byuons does not depend on i and may take by magnitude only two values:

$$
\begin{array}{rcl}
X_{II^- I^+}^{k-i,i} = \frac{const_1}{A_G};  & \quad &
X_{II^+ I^-}^{i+1,k-i-1} = -\frac{const_1}{A_G};\\
X_{I^+ II^+}^{i,i+2} = \frac{const_1}{k \cdot A_G};  & \quad &
X_{I^- II^-}^{k-i-2,k-i} = -\frac{const_1}{k \cdot A_G};\\
\end{array}
$$

   or multiples of them, for example,

$$
X_{II^- I^+}^{Nk-i,i} = \frac{const_1 N}{A_G} .
$$

The expressions determining the maximum energy of byuon interaction are written as

\begin{equation}
\begin{array}{rcl}
E_{II^- I^+}^{k-i, i} = \frac{[A x]_{II^-}^{k-i}[A x]_{I^+}^{i}}{X_{II^- I^+}^{k-I,i}} & = & \frac{const_1}{4 k^2} A_G [2(k-i) - 1] (2i + 1);\\
E_{II^+ I^-}^{i+2,k-i-2} & = & \frac{const_1}{4 k^2} A_G [2(k-i) - 1] (2i + 1);\\
E_{I^+ II^+}^{i,i+2} & = & -\sqrt{-1}\frac{const_1}{4 k^2} A_G  (2i + 3) (2i+1);\\
E_{I^- II^-}^{k-i-2,k-i} & = & \sqrt{-1} \frac{const_1}{4 k^2} A_G [2(k-i) - 1] [2(k-i) -3];\\
E_{II^- II^-}^{NkP-i-k, NkP-i} & = & \frac{const_1}{4 k^2} A_G [2(NkP-i-k) - 1] [2(NkP-i) -1].\\
\end{array}
\end{equation}

   The minimization of $\Delta E(i)$ is achieved in the functional space of the following variables:
\begin{equation}
\Psi_{I^+}^0, \, \Psi_{II^+}^2, \, \Psi_{II^-}^{NkP}, \, \Psi_{I^-}^{NkP-2}, \cos_{I^+ II^+}^{i,i+2}, \, \cos_{II^- II^-}^{NkP-k-i,NkP-i}, \, k, \, NP.
   \end{equation}

It is assumed therewith that the conditions of symmetry during the
" closure of the loop" in i are fulfilled as well as symmetry of
the world and antiworld, which conditions can be represented as:
\begin{equation}
\begin{array}{rcl}
\cos_{I^+ II^+}^{i,i+2} & = & \cos_{I^- II^-}^{NkP-i-2,NkP-i} = \cos_{I^- II^-}^{NkP-i-2-2,NkP-i-2}\\
& = & \cos_{I^- II^-}^{NkP-i-2-2 \times 2,NkP-i-2 \times 2} = \ldots = cos_{II^- II^-}^{NkP-i-k,NkP-i}\\
& = & \cos_{II^- II^-}^{NkP-i-2k,NkP-i} = \ldots
\end{array}
\end{equation}

The functions $\cos_{II^+ I^-}$, $\cos_{II^+ I^+}$ are considered
as equal to 1.

Initial conditions for $\Psi$- function are preset to be
\begin{equation}
\begin{array}{rcl}
\Psi_{II^+}^0 \approx 0, \, \Psi_{I^+}^0 \approx 1, \, & \Psi_{II^-}^{NkP} \approx 1, & \Psi_{I^-}^{NkP} \approx 0, \, \Psi_{I^+}^i \approx \Psi_{I^-}^{NkP-2}\\
\Psi_{II^-}^{NkP-1} = \Psi_{II^+}^2 & & \Psi_{I^-}^{i+2} + \Psi_{I^-}^{i+1} = 0.
\end{array}
\end{equation}

Using the solutions of the equations (8) and (9):
$\Psi_{II^+}^{i+1} + \Psi_{I^+}^{i+1} = A \cos \left( \frac{2 \pi
i}{k} \right) + B \sin \left( \frac{2 \pi i}{k} \right)$ we
contract the space of variables down to four: $\cos_{I^+
II^+}^{i,i+2},$ $\cos_{II^- II^-}^{NkP-i-2,NkP-i},$ $k, NP$.

Now, taking into account the normalizing expressions (5), seek for min $\Delta E(i)$ by the steepest descent method. When retaining only 14 terms of the series in (10) and (5a), min $\Delta E(i)$ will correspond to the following values:
\begin{equation}
\begin{array}{rcl}
\Psi_{I^+}^0 = 0.999(6), & \Psi_{II^+}^2 = 1.00136 \times 10^{-4}, & \Psi_{II^-}^{NkP} = 0.999(8)\\
\Psi_{I^-}^{NkP-2} = 1.100043 \times 10^{-4}, & \cos_{I^+ II^+}^{i, i+2} & = 1.01887 \times 10^{-5},\\
\cos_{II^- II^-}^{NkP-i-k,NkP-i} = 1.20013 \times 10^{-5},& k = 6.2 \times 10^{15}, & NkP = 3 \times 10^{60}\\
\end{array}
\end{equation}

With increasing n, as is seen from the solutions given and Fig.6, $k$ (the first period in $i$) approaches its value obtained in Refs. [25-26] on the base of physical considerations as an integer part of the ratio $x_0 / \tilde x_0 = 3.2 \times 10^{15}$. Thus, we can now obtain, with the aid of the calculated $k$, one of the fundamental dimensions in physics of elementary particles, $x_0 \approx 10^{-17} \, cm$, with the only quantum of space $\tilde x_0$ given. This mathematical result raises prospects that the advanced hypothesis is true. It reflects the nature of physical space and vacuum.

The minimum $\Delta_1E(i)$ was sought for a case when $Nk < i \le NkP$. In this case the normalizing expressions for the arising interactions of byuons have the form:
$$
\sum_{j=Nk}^{j=1} \sum_{\xi=0}^{\xi=\frac{NkP-i-2}{2}} \Psi_{I^+}^j \Psi_{II^+}^{j+2} \Psi_{I^-}^{NkP-j-2-2\xi} \Psi_{II^-}^{NkP-j-2\xi} =\frac{i N P}{2 k};
$$

\begin{equation}
\begin{array}{rcl}
\sum_{\xi=1}^{\xi=\frac{NkP-i-2}{2}} \Psi_{II^-}^{NkP-i} \Psi_{II^-}^{NkP-i-\xi k} & =& \frac{ N k P - i}{N k};\\
\sum_{\eta=1}^{1/k} \Psi_{II^+}^{i+2} \Psi_{II^+}^{i+2-\eta k} & =& \frac{i}{N k};\\
\sum_{\xi=0}^{\xi=\frac{NkP-i-2}{2}} \sum_{j=Nk}^{j=1}
\Psi_{II^+}^{j+2} \Psi_{I^-}^{NkP-j-2-2\xi} & =& \frac{ i N P}{2 k};\\
\sum_{\xi=0}^{\xi=\frac{NkP-i-2}{2}} \sum_{j=Nk}^{j=1} \Psi_{II^-}^{NkP-j-2\xi} \Psi_{I^+}^{j} & =& \frac{i N P}{2 k}.\\
\end{array}
\end{equation}

The expression for $\Delta_1 E(i)$ becomes more complicated:

$$
\begin{array}{rcl}
\Delta E_{1} (i) & = & \Psi_{II^+}^{i+2} \cdot \Psi_{II^+}^{i+2-k} \cdot E_{II^+ II^+}^{i+2, i+2-k } \cos_{II^+ II^+}\\
& + & \Psi_{II^+}^{i+2} \cdot \Psi_{II^+}^{i+2-2k} \cdot E_{II^+ II^+}^{i+2, i+2-2k } \cdot \cos_{II^+ II^+} + \ldots\\
& + & \Psi_{II^+}^{i+2} \cdot \Psi_{I^-}^{NkP-i-2} E_{II^+
I^-}^{i+2,NkP-i-2} \cdot \cos_{II^+ I^-} +\\
& + & \Psi_{II^+}^{i+2} \cdot \Psi_{I^-}^{NkP-i-2-2} E_{II^+ I^-}^{i+2,NkP-i-2-2} \cdot \cos_{II^+ I^-} +\\
& + & \Psi_{II^+}^{i+2} \cdot \Psi_{I^-}^{NkP-i-2-2 \times 2} E_{II^+ I^-}^{i+2,NkP-i-2-2 \times 2} \cdot \cos_{II^+ I^-} +\ldots +\\
& + & \Psi_{II^+}^{i+2-k} \cdot \Psi_{I^-}^{NkP-i-2+k} E_{II^+ I^-}^{i+2-k,NkP-i-2+k} \cdot \cos_{II^+ I^-} +\\
& + & \Psi_{II^+}^{i+2-k} \cdot \Psi_{I^-}^{NkP-i-2-2+k} E_{II^+ I^-}^{i+2-k,NkP-i-2-2+k} \cdot \cos_{II^+ I^-} +\\
& + & \Psi_{II^+}^{i+2-k} \cdot \Psi_{I^-}^{NkP-i-2-2 \times 2+k} E_{II^+ I^-}^{i+2-k,NkP-i-2-2 \times 2+k} \cdot \cos_{II^+ I^-} +\ldots +\\
& + & \Psi_{II^+}^{i+2-2k} \cdot \Psi_{I^-}^{NkP-i-2+2k} E_{II^+ I^-}^{i+2-2k,NkP-i-2+2k} \cdot \cos_{II^+ I^-} +\ldots + \Delta E(i)\\
\end{array}
$$

($\Delta E(i)$ is taken from Eq.10),  taking into account for multiplication over index i as was shown above ($i+2, \, i+2-k, \, i+2-2k$  etc.).

The search for min $\Delta E(i)$ with the use of the chain of equations (7) in the space of similar variables with similar initial conditions (where $i = 0$ corresponds now to $i = Nk$), leads to practically the same results:

$$
\begin{array}{rcl}
\Psi_{I^+}^{Nk} = 0.9998,  \Psi_{II^+}^{Nk+2} = 1.006 \times 10^{-4}, \Psi_{II^-}^{NkP} = 0.9999\\
\Psi_{I^-}^{NkP-2} = 1.101 \times 10^{-4},  \cos_{I^+ II^+}^{i, i+2}  = \cos_{II^+ II^+}^{i+2, i+2-k} = 1.1019 \times 10^{-5},\\
\cos_{II^- II^-}^{NkP-i-k,NkP-i} = 1.203 \times 10^{-5}, k = 4.475 \times 10^{15},  NkP = 3.169 \times 10^{60}\\
\end{array}
$$

It is interesting that at $i < k$ and even at $i \approx Nk$ (Table 1), a significant part in the magnitude of $\Delta E(i)$ is potential energy of the byuon interaction in $VS \, II^- II^-$, and at $i \rightarrow NkP$ the four-contact interaction of byuons in $VS \, II^+ I^+ I^- II^-$  becomes determining.

\begin{table}
\caption{ }
\begin{tabular}{ccc}
\hline
E [erg] \, i & Nk & NkP\\
$E_{II^+ II^+}$ & $10^{13}$ & $10^{95}$\\
$E_{II^+ I^+ I^- II^-}$ & $10^{70}$ & $10^{111}$\\
$E_{II^- II^-}$ & $10^{95}$ & $10^{13}$\\
\hline
\end{tabular}
\end{table}

Let us consider a simplified case when only the period in $i$,
equal to $k$, is present in the antiworld. Then we have from the
necessary condition of minimum $\Delta E(i)$ with respect to the
function $\Psi_{II^+}^{j+2}$ (i.e. from the equation
$\frac{\partial \Delta E(i)}{\partial \Psi_{II^+}^{j+2}} = 0$) we
have
\begin{equation}
\cos _{II^+ I^+} = -\frac{1}{\Psi_{I^+}^{i} \cdot
\Psi_{II^+}^{k-i}} \cdot \frac{1}{k} \cdot \frac{\sqrt{[(k-i) - 1]
(2i+1)}}{(2i+3) [(2k-i) + 3]} .
   \end{equation}

If $i \ll k, \, N=P=1$, and hence, according to Eq. (5)
$\sum_{j=0}^{j=1} \Psi_{I^+}^{j} \cdot \Psi_{II^-}^{k-j}
\rightarrow 1$, we obtain from Eq.(17) $\cos _{II^+ I^+} \approx
1/k$.

Thus, as show our numerical calculations and analytical estimations, the minimization of $\Delta E(i)$ leads to values of functions $\cos _{II^+ I^+}$ etc. if not zero but extremely small.

Note that the functions $\cos _{II^+ I^+}$   for the case of $NP$
"loops" in $VS \, II^-$  and for the case when $N=P=1$, differ by
ten orders of magnitude. The physical meaning of these functions
will be shown further.

Put a question, where disappears and into which transforms the
potential energy of byuon interaction? An answer seems to be
simple, of course,  into the kinetic energy of rotation (since
$\cos _{II^+ I^+} \ll 1$, see (15), (17)). But the rotation of
what and about what? And why do we assume the law of conservation
of energy to be fulfilled here? After all, it makes no sense to
say about uniformity in time for this statement in which the time
is discrete! Let us answer these questions.

As it was shown above, the optimum values of the functions $\cos
_{II^+ I^+}^{i+2,i}$, $\cos _{II^- I^-}^{NkP-I, NkP-i-2}$,  etc.,
are much less than 1 but non-zero. The smallest values of $\cos
_{II^+ I^+} \approx 1/k$ correspond to residual (finite) potential
energies $E_k$, from which, as we will further assume, the
smallest part (associated with the formation of the own space of
elementary particles) of the potential self-energy of elementary
particles corresponding to the known Einstein's relationship $E_k
= m c^2$, is added together. Determine the minimum value of $E_k$.
It is seen from Eq.(10) that for the simplest objects with $N$ and
$P$ approximately equal to 1, the minimum $E_k$ is equal to the
potential energy of the four-contact interaction of byuons with
the minimum values $\cos _{II^+ I^+}, \,\cos _{II^- I^-}$.

In view of the normalization (5) we have for this case
\begin{equation}
E_{k_{min}}^{(0)} = \sqrt{E_{ I^+ II^+}^{0,2} \cos _{I^+ II^+} E_{ I^- II^-}^{k-2,k} \cos _{I^- II^-}} .
\end{equation}

From here, with the condition (13), and using Eqs.(11) and the equality $\cos _{II^+ I^+} \approx 1/k$, we obtain $ E_{k_{min}}^{(0)} \approx 33 \, eV$.

Consider the process of energy transformation for the four-contact
byuon interaction occurring within a time quantum $\tau_0$. Hence,
the transformation must be discrete.

Any value of i can be re-denoted with an another index, for
example, with $j, \, \gamma, \, \xi$, etc., which may be set equal
to zero. At each point then, where $j = \gamma = \xi = \ldots =
0$, there will be always present an own system of account of the
indices $j, \, \gamma, \, \xi$, etc., as well as the minimum
energy of four-contact byuon interaction $ E_{k_{min}}^{(0)}
\approx 33 \, eV$. Note that this minimum energy is limited in
index $i$ by values $i = 0$  and $i = 2$ (point A in Fig.3).

Thus, we have in $R_1$ two sets of points $A(\{A\})$ and
$D(\{D\})$, between which  the dynamic process of renumeration
goes in connection with the properties of vacuum states of the
discrete objects, the byuons $II^+, \, I^+, \, I^-, \, II^-$:

$$
\begin{array}{rcl}
the \, time \, \tau_{i+1} & &
\begin{cases}
\{D\} = i + 1, \, i - 1, \ldots\\
\{A\} = i + 2, \, i, \ldots\\
\end{cases}
\\
the \, time \, \tau_{i+1} + \tau_0 & &
\begin{cases}
\{D\} = i + 2, \, i, \ldots\\
\{A\} = i + 3, \, i+1, \ldots\\
\end{cases}
\\
\end{array}
$$

The set $R_1$ may be represented, at some $i$-th point of time as the join of $\{A\}$ and $\{D\}$, i.e. $R_1 = \{A\} U\{D\}$ (Fig.7).

Hence the space $R_1$ segregates into the subspaces $R_D$ of
$D$-points and $R_A$ of $A$-points. Thus, we may say about the
motion of $D$-points relative to $A$-points. A new, second
coordinate is appeared, symbolize it by $Y_{AD}$ (Fig.7). The
minimum value of $Y_{AD}$ is
 $Y_{AD_{min}} = C_0 \tau_0 |(i+1)_A- iD|
 = \tilde X_0$.  The minimum plane object appears.Assume that its
appearance (new coordinate) corresponds to the minimum action $h$
( see [20-22],  $h = (([A_{g} x_0]_{II^+} [A_{g} x_0]_{I^-}) /
c_0) x_0 / c t^*$  and elementary electric charge $e_0^2 = 1 / (4
\sqrt{3})) A_g^2 x_0^2 (x_0) / c t^*)^{3/2}$  We may then
introduce the concept of momentum for objects with the residual PE
of byuon interaction by writing the relationship $P_{D}^{i+1}
\times Y_{AD}^{i,i+1} = h$ where $P_{D}^{i+1}$  is the momentum of
the point $D$ numbered $i+1$ relative to the point $A$ with the
number $i$. Similar relationship can be written in any point from
$\{A\}$. The direction of the momentum vector $\bar P_D^{i+1}$
corresponds to that towards the point $D^{i+3}$ of the subspace
$R_D$. The direction of the coordinate corresponds to the vector
directed from the point $A^i$ to the point $D^{i+1}$.

The appearance of the minimum plane object and realization of the
minimum action are connected with the origin of the quantum spin
number $\bar S = [\bar P_D^{i+1} \times \bar Y_{AD}^{i,i+1}]$
expressed numerically in minimum actions $h$.

The functions $\cos _{II^+ I^+}$ etc. minimizing $\Delta E(i)$,
will be further considered by us as cosines of the angles between
the vector $\bar P_D^{i+1}$ and $\bar Y_{AD}^{i,i+1}$. I.e. before
the byuon interaction, the space $R_1$ represents, at some time
point $\tau$, a certain discrete straight line of points $\{D\}$
and $\{A\}$, and at the time point $\tau + \tau_0$ forms a line,
broken at points $\{D\}$. That is the minimum interaction of
byuons has occurred.

If an "observer" was able to perceive objects with $E > 0$ only
every $\tau = k \tau_0$, he would simultaneously (within a time
quantum $\tau_0$) fix all the planes arrived to the point of
"observation", and "see" already the three-dimensional world
formed from the plane world in the result of its dynamics within
the time $\tau = \tau_0$. Why the three-dimensional and not
N-dimensional one? Because the set of two-contact interactions of
byuons is divided, depending on reference point (Fig.2, $i = 0, \,
1, \, 2$), into three subsets $M_0, \, M_1, \, M_2$ (the lower
index denotes reference point), corresponded by three
one-dimensional subspaces $R_{1,0}, \, R_{1,1}, \, R_{1,2}$ while
introducing metric properties. Explain the above said.

As was indicated above, any value of index i can be always redenoted by j and then $j = 0, \, 1, \,2$ corresponds to reference points. Redenoting $i+1$ by $\xi$, $i+2$ by $\gamma$ etc. leads, depending on reference points, to formation of three families of subspaces embedded in each other:

   reference point "0" $\Rightarrow R_{1,0}^i \supset R_{1,0}^j \supset R_{1,0}^{\xi} \supset R_{1,0}^{\gamma}$ etc.;

   reference point "1" $\Rightarrow R_{1,1}^i \supset R_{1,1}^j \supset R_{1,1}^{\xi} \supset R_{1,1}^{\gamma}$ etc.;

   reference point "2" $\Rightarrow R_{1,2}^i \supset R_{1,2}^j \supset R_{1,2}^{\xi} \supset R_{1,2}^{\gamma}$ etc., \quad if $i > j > \xi > \gamma$  etc.

Thus, in connection with the existence of three independent
reference points for the new pair interactions, the three
independent coordinates should be given to fix a pair interaction
with respect of the three reference points, i.e. $R_3$ can be
represented as $R_3 = R_{1,0} \times R_{1,1} \times R_{1,2}$. Note
that $R_{1,0}, \, R_{1,1}, \, R_{1,2}$  consist of the sets of
points $\{A\}$ and $\{D\}$, i.e. at each subsequent point in time,
renumeration of points $A$ and $D$ and "spinning" of objects with
$E > 0$  in the subspaces $R_{1,0}, \, R_{1,1}, \, R_{1,2}$,
occur. In this manner the concept of spin is introduced for
objects with $E > 0$ in $R_3$. For objects of great size (as a
result of described minimization of PE), the rotation will be
always take place since the byuons are not closed in a volume of
$R_3$ due to $\cos _{II^+ II^+} \ne 0$. That is why, planets,
stars and so on rotate, with the main part of potential energy of
byuons being transformed into energy of rotation. The latter is
because the functions $\cos _{II^+ II^+}$, $\cos _{I^+ II^+}$ etc.
are very small.

Advance (without proving) the following theorem:

\textbf{If a system is closed, the amount of information in it is
constant.}

That is, transformation of one information image into another is possible in the system, but the total amount of information does remain invariable.

Note that by information we mean here not informativity as in
theory of information developed by Hartly and Shannon [29] on the
basis of entropic approach, but the numbers of information bits
(the values "0" and "1") in one or another information subsystems
of the system of considered objects (the combinatoric approach
[30]). By "1" we imply here accomplishing the minimal act (minimum
action $h / 2$) in the system with formation of an object with $E
> 0$ from byuons, and by "0" disappearing of the object with $E >
0$ is meant.

On the base of the theorem, write the following equality for informational units (bits) in the subspaces $R_1$ and $R_3$:
\begin{equation}
2^{\frac{\tau_0 E_1 S_1 k \cdot 2}{h}} = 2^{S_3 N}
\end{equation}

   where $N$ is the number of information images in $R_3$ ($N = c t^*/x_0$ - the second period in $i$);  $S_3$ the complexity of the information image in $R_3$ (number of "loops" of length $N$; $S_3 = 1, \, 2, \ldots$); $E_1$ potential energy of minimum four-contact interaction of byuons in $R_1$  if $\cos _{II^+ I^+} = \cos _{II^- I^-} = 1$; $\frac{\tau_0 E_1 S_1 k \cdot 2}{h}$ is a transformation factor of recounting the number of information images ($k$) in $R_1$ into that in $R_3$ for $i \ge k N$; $S_1$ is complexity of the information image in $R_1$ (number of "loops" of length $k$, $S_1 = 1, \, 2, \ldots$).

 By an information image in $R_3$, one means one or another quantum number of an elementary particle.

With Eq.(19), the expressions for lepton masses obtained in
[20-22]( see Appendix 1] become more understandable (for example,
$m_e c^2 = E_{k_{min}}^0 \cdot N = 33 \, eV \cdot c t^* / x_0 = 2
m_{\nu_e} c^2 \cdot c t^* / x_0$).

Taking $k$ from the solution of the problem of searching minimum
$\Delta E$  and substituting $S_3 = 1, \, S_1 = 1$ into Eq.(19),
we find $N = 1.544 \times 10^4 = c t^*/x_0$, and knowing $N$ and
$N P$ from this same solution, determine $P \approx 10^{42}$, and
$\tilde x_0 k N \approx 10^{-13} \, cm$.

Thus, we find all the periods of byuon motion in $i$, corresponding to the following scales of our World: $10^{-17} \, cm$ is a characteristic scale of weak interactions (for larger lengths our World is three-dimensional and almost orthogonal, for an empty space with the $10^{-15}$ precision); $10^{-13} \, cm$ is a characteristic size of proton and atomic nuclei), $10^{28} \, cm$ is the radius of our Meta-galaxy or the observable part of our Universe.

 (See: Appendix 1 -- The expressions for masses of ultimate particles; Appendix 2 - A qualitative distinction between the theory of byuons and previous physical theories. Appendix 3 - Force-free physics. A qualitative pattern of a common approach to unifying all interactions. A novel principle of relativity.)

It should be also noted that to calculate the fundamental
constants $h, \, e_0, \, c$; constants of known interactions,
masses of main baryons, leptons, and mesons , only three numbers
$\tilde x_0, \, \tau_0, \, |\mathbf{A}_g|$ should be given since
the characteristic dimensions $\tilde x_0 \approx 10^{-17} \, cm,
\, c t^* \approx 10^{-13} \, cm$ and $10^{28} \, cm$ are found
from the minimum PE of byuons and from the information theorem.

\section{Dark Matter and Dark Energy in the byuon theory}

Determine the average density of substance in the Universe while
taking $i = N k P$ and, hence, its characteristic dimension
$\tilde x_0 N k P \approx 10^{28} \, cm$ (it coincides with the
assumed radius of the Universe). The total energy in the Universe
can be represented as $\frac{h}{\tau_0} N k P$. Its value is $5.4
\times 10^{77} \, erg$, and the corresponding equivalent mass
$\approx 6 \times 10^{56} \, g$. The uniformity of distribution of
substance over the sphere with the radius $\tilde x_0 N k P$ gives
the density of substance in the Universe $\approx 10^{-29} \,
g\cdot cm^{-3}$, which is measured in the known observations.

Through the set $\{A\}$, information exchange occurs between points of the set $\{D\}$, which is the main mechanism determining physical essence of Heisenberg uncertainty interval in the conception of physical space and physical vacuum. Without introducing the points $\{A\}$, the connection between $D$-points is realized in one direction only, in that of increasing index $i$ with the speed no greater than $c_0$.

According to the developed conception of physical vacuum
structure, we can determine momentum and a coordinate of such a
complex object as an elementary particle only with an accuracy of
the momentum and coordinate entering into the relationship
$P_{D}^{i+1} Y_{AD}^{i,i+1} = h$ governing the momentum of
$D$-points mapped into $R_3$ according to (19). That is, writing
the uncertainty relation in $R_3$ for some elementary object
$\gamma$ as $\Delta P \cdot \Delta X \le h/2$, we mean $\Delta P$
and $\Delta X$ to be caused by the process of $R_3$ formation from
byuons, i.e. determined by quantities of $P_D$ and $Y_{AD}$ type.

Its momentum corresponding to the minimum momentum for elementary particles, can be given in general form as [20-22]

$$
P = \Phi \cdot E_{k_{min}}^0 / c,
$$

where $\Phi$ is probability of observing the object $4 b$ in some region of space $R_3$.

If the objects $4 b$ are free (that is, they create not an elementary particle but space free of them), then $\Phi = \frac{1}{16} \cdot \frac{\tilde x_0^3}{4 \pi x_0^2 \tilde x_0}$, where $\tilde x_0 \approx 10^{-33} \, cm$, and $x_0 \approx 10^{-17} \, cm$. In this case, if the scatter in values of momentum is $P$ for an elementary object $\Delta p$, then the uncertainty in coordinate in $R_3$ for the object $4 b$ will be equal to $10^{28} \, cm$. This value $\Delta X$ has given us earlier the possibility to obtain the density of matter in the Universe, observed in experiment, by way of averaging it over the sphere $10^{28} \, cm$ in radius.

If the object $4 b$ is not free (that is, it forms the internal geometry of an electron, for instance), then

\begin{equation}
\Phi = \frac{1}{16} \cdot \frac{x_0^3}{4 \pi (N x_0)^2 x_0} ,
\end{equation}

and we can write the following expression for an assemblage of
objects $4 b$ forming an electron (for which $m_e c^2 = N \cdot
E_{k_ {min}}$):

\begin{equation}
\Delta p = \frac{1}{16} \cdot \frac{x_0^3}{4 \pi (N x_0)^2 x_0}
\cdot \frac{N E_{k_{min}^0}}{c} = \frac{1}{64 \pi} \cdot \frac{N
E_{k_{min}^0}}{N c}.
\end{equation}

Using Eq.(21) we obtain the uncertainty in coordinate $\Delta x$ in $R_3$ on the order of $10 \, cm$ for the assemblage of $N$ objects $4 b$, that is, the electron, due to wave properties of $N$ objects $4 b$, carries information on its properties not over distances of $10^{-8} \, cm$ (characteristical dimension of de Broglie wave for electron at the temperature of $300 K$) as would be in the case of a pointwise particle but over distances of the order of $10 \, cm$.

When considering not $N$ objects but one object $4 b$ in the electron (that is, when Formula (20) is valid) then $\Delta x \approx 10^5 \, cm$. Hence the less is amount of information on condition of internal spatial characteristics of electron, the larger is the scatter in coordinate.

In Appendix 3, qualitative pattern of a common approach to
unifying all interactions is shown

The byuon theory predicts a new anisotropic interaction of nature objects with physical vacuum.

Peculiar "taps" to gain new energy are elementary particles
because their masses are proportional to the modulus of some
summary potential $\mathbf{A}_{\Sigma}$ that contains potentials
of all known fields (Appendix 1). The value of
$\mathbf{A}_{\Sigma}$ cannot be larger than the modulus of
$\mathbf{A}_g$ [20-22]. In accordance with the experimental
results shown in [22,31], this force ejects substance from the
area of the weakened $\mathbf{A}_{\Sigma}$ potential along cone's
formations with solution $100^{\circ} \pm 10^{\circ}$ formed
around vector $\mathbf{A}_{\Sigma}$ direction. This vector has the
following coordinates in the second equatorial system of
coordinates: right ascension $\alpha \approx 293^{\circ} \pm
10^{\circ}$ ($19^h 20^m$), declination $\delta \approx 36^{\circ}
\pm 10^{\circ}$ [22,31]. The vector $\mathbf{A}_{\Sigma}$  is
parallel to the vector $\mathbf{A}_g$  practically.

The new force is of nonlinear and non-local character as to
variation of some summary potential $\mathbf{A}_{\Sigma}$ and may
be represented by some series in $\Delta \mathbf{A}_{\Sigma}$
[20-22,32].

The expression for the new force takes the form :

\begin{equation}
| \textbf{F}| = -2 N m_0  A_g \cdot c^2 \lambda (\Delta A) \cdot
\frac{\partial \lambda (\Delta A)} {\partial \Delta A} \cdot
\frac{\partial \Delta A} {\partial X_1} ,
\end{equation}

where $N$ is the number of stable elementary particles in the
body, $A_g = |\textbf{A}_g|$.

Note that expression for the new force (22) is local (we cannot
deal with the nonlocal ones as yet), therefore, to account for the
nonlocality of the phenomenon, we will take $\Delta A$ equal to
the difference in changes of the summary potential $|
\textbf{A}_{\Sigma}| $ at the location points of a test body  and
a sensor element [20-22].

These changes being equal, the force will be absent. Depending on the relative position of the sensor and the test body, $\Delta A$ can take as a positive, so a negative value.

To estimate a role of gravitational field in a change of
$|\textbf{A}_{\Sigma}|$ we put forward the maximal gravitational
potential $\varphi_{max}$,   determined for proton by the
following relationship:

\begin{subequations}

\begin{equation}
m_p \varphi_{max} = e A_g ,
\end{equation}

where $m_p$ is the proton mass. Then  the contribution of
$\varphi_{max}$ in the change of $|\textbf{A}_{\Sigma}|$ is
described by the following equality:

\begin{equation}
\varphi_{max} \cos\alpha \cdot m_p = e A v / c,
\end{equation}
\end{subequations}

where $v$ is in this case the velocity of our Galaxy relative to
the neighbouring galaxies, $\cos\alpha = \cos_{II^-
II^-}^{NkP-i-2, NkP-i}$ (15). It characterizes nonorthogonality of
our World at the moment of the formation of the space of
elementary particles. It is worth noting that potentials of
physical fields have the physical meaning only for interacting
byuons when elementary particles are generated with their masses
and charge numbers.  As for the vector potential of magnetic field
it is gauged   so that its value on the axis for example of the
solenoid equals zero; $x_1$ is coordinate, directed from the point
of the most decreased $ \textbf{A}_{\Sigma}$ on a winding to the
vector $ \textbf{A}_g$.

The analysis of the specific experimental results with high field magnets (see [20-22,32-35]) has led to the following expression for $\lambda (\Delta A)$:
\begin{equation}
\lambda (\Delta A) = \sum_{k=1}^{\infty} \lambda_k \exp \left\{ -
\left[ \frac{\Delta A}{A_g} \cdot \frac{r}{\Delta y} \cdot \left(
\frac{c t^*}{x_0} \right)^{3/2} \right]^k \right\} \cdot \Delta
A^k ,
\end{equation}

where $r$ is the radius of the circle which the test body is
located on; $\Delta y$ is the difference in coordinates $y$ of the
sensor and the body [20-22]; $(x_0 / c t^*)^{3/2}$ is part of the
energy $2 m_{\nu_e} c_0^2 = 2 m_0 |A_g| c_0^2$, which can be acted
upon by the electromagnetic field potentials.

Using the linear term only in the expansion  of (22) by $\Delta A$, we obtain the following formula for the modulus of the new force:

\begin{equation}
F =2 N m_{\nu} c^2 \lambda_1^2 \cdot \Delta A_{\Sigma} (\Delta A_{\Sigma} / \Delta X).
\end{equation}

It is worth noting that the experiments for the scanning the celestial sphere by the pulsed plasma generator  [22,31] to detect some directions in space, where energy is  more than the average value, are the final stage in the determination of the direction of the new force. This direction was determined before by the using the high field magnets [20-22,32-35], by the investigations of the rate of $\beta$ - decay for a number of radioactive elements [20-22,24,25,36], by the investigations with high precision gravimeters  [20-22,37], and by plasma generator of other types  [20-22,38].

Experiments with high field magnets [20,21] showed that the new
interaction has the most probably an isotropic component as well.

Let us discuss the nature of the dark energy in the framework of the byuon theory on the base of potentials of physical fields.

It is known that the gravitational potential $\varphi$ is
negative, and therefore for any summation of potentials it
decreases the modulus of $\mathbf{A_{\Sigma}}$. Masses of
elementary particles are proportional to this modulus. Hence the
new force will push out any material body from the region of the
decreased modulus of $\mathbf{A_{\Sigma}}$, because  a defect of
energy $\Delta E = \Delta m c^2$ will appear and the corresponding
force will act to the region with undisturbed value of
$\mathbf{A_{\Sigma}}$.  Any material body  decreases in its own
region the modulus $\mathbf{A_{\Sigma}}$ due to potentials of
physical fields of all its elementary components, i.e. creates the
gradient $\Delta A_{\Sigma} / \Delta X$. Gravitationally acting
mass, for example,  our Galaxy, creates around the gravitational
potential $\varphi$. To estimate the action of one galaxy to
another we put in the formula (25) the potential  $\varphi_{max}$
from (23a) and  $\Delta A_{\Sigma}$ from (23b) ($\Delta A_{\Sigma}
= A$). Let us estimate  the distance $R_{GG}$ where the new force
$F$ from (25) will be higher than the gravitational force $F_g$
(cf. Fig.1):

\begin{equation}
R_{GG} \ge G M_g^2 / (2 N m_{\nu} c^2 \lambda_1^2 \cos^2\alpha
\varphi_{max}^2 (m_p c / v e)^2)
\end{equation}

where $G$ is the gravitational constant.

Here  ${M_g}$  is the mass of the one of interacting galaxies.  We
consider an interaction of two galaxies with $10^{10}$ stars,
assume that  the mass of each star is of order to the solar mass
($\sim 10^{33} \, g$) and  a relative  velocity of each galaxy $v
= 100 \, km/sec$ and $1000 \, km/sec$. From our experiments
$\lambda_1 = 10^{-12}$ [20-22,51]. As the result  we obtain from
(26)  $R_{GG} \ge 10^{26} \, cm$ for $v = 100 \, km/sec$ and
$R_{GG} \ge 10^{28} \, cm$ for $v = 1000 \, km/sec$.

Thus we have estimated the magnitude of the distance between galaxies above which they scatter under the action of the new force. The estimate obtained seems as reasonable and indicates that the physics of byuons is perspective to explicate the nature of the dark energy and dark matter.

\appendix
\section*{Appendix 1.}

The masses of ultimate particles can be described by the following
expressions [20-22]:

\begin{equation}
\begin{array}{rcl}
m_p c^2 & = & \frac{\sqrt{h c}}{2} \left( \frac{4 \sqrt{3}
e_0^2}{h c} \right)^3 \left(1 - \frac{1}{2} \sqrt{\frac{3 x_0}{c
t^*}} \right) \times \left[ \sqrt{1 + \frac{1}{6 \left( 1 -
\frac{1}{2} \sqrt{\frac{3 x_0}{c t^*}} \right)^2}} + 1 \right]
|{\textbf{A}}| = 924 MeV,
\\
m_{\pi^0} c^2 & = & \frac{\sqrt{h c}}{2} \left( \frac{4 \sqrt{3}
e_0^2}{h c} \right)^3 \left(1 -  2 \sqrt{\frac{3 x_0}{c t^*}}
\right) \times \left[ \sqrt{1 + \frac{1}{6 \left( 1 - 2
\sqrt{\frac{3 x_0}{c t^*}} \right)^2}} -1 \right] |{\textbf{A}}| =
132 MeV,
\\
\end{array}
\end{equation}

   Masses of all leptons:

\begin{equation}
\begin{array}{rcl}
2 m_{\nu_0} c^2 = \frac{\sqrt{h c}}{2 k} \cdot \left( \frac{h c}{4
\sqrt{3} e_0^2} \right) \cdot |{\textbf{A}}| , & & m_{e} c^2 =
\frac{\sqrt{h c}}{2 k} \cdot \left( \frac{h c}{4 \sqrt{3} e_0^2}
\right) ^3 \cdot |{\textbf{A}}| ,\\ 2 m_{\nu_{\mu}} c^2 = 2 \cdot
3^{1/4} \sqrt{h c} \cdot \left( \frac{4 \sqrt{3} e_0^2}{h c}
\right) ^{5.5} \cdot |{\textbf{A}}| , & & m_{\mu} c^2 = 3^{1/4}
\sqrt{h c} \cdot \left( \frac{4 \sqrt{3} e_0^2}{h c} \right)
^{3.5} \cdot |{\textbf{A}}| ,\\ 2 m_{\nu_{\tau}} c^2 = 3^{1/4}
\cdot 96 \cdot \sqrt{2} \cdot \sqrt{h c} \cdot \left( \frac{4
\sqrt{3} e_0^2}{h c} \right) ^{5.5} \cdot |{\textbf{A}}| ,
&  & m_{\tau} c^2 = 3^{1/4} \cdot 12 \cdot \sqrt{2} \cdot \sqrt{h c} \cdot \left( \frac{4 \sqrt{3} e_0^2}{h c} \right) ^{3.5} \cdot |{\textbf{A}}| .\\
\end{array}
\end{equation}

\section*{Appendix 2.}

It is known that any novel physical model of the Universe must
meet the following criteria. First, all the discovered laws of
nature as well as sufficiently well established models of one or
another physical phenomena must follow from the new model as
asymptotical approximations. Second, the new theory should have
the capability for predictions. That is, it should guide an
experimental way to the gain of new knowledge, as the theory
itself gives nothing but only points such a way. Criterion of
truth is an accurately performed experiment independently
confirmed by various authors. The theory of byuons [20-22] seems
to meet the above criteria. That is a theory of "life" of special
discrete objects from which the surrounding space and the world of
elementary particles form. The intrinsic dynamics of byuons
determines such fundamental phenomena as the course of time,
rotation of planets and stars, spins of elementary particles, asf.

What is a qualitative distinction between the theory of byuons and previous physical theories?

First, the physical space was always given in all the science, in
one way or another, and motion equations for a system of objects
under study were written in that space. Space could be uniform
continuum (Newton, Minkovsky) or discrete, one-dimensional or
multidimensional, asf. In present-day cosmological models of the
Universe origin (Gamov's Big Bang [39], Linde's model of bulging
Universe [40], and so on), space is always given, too. But in the
theory of byuons, the physical space (necessarily
three-dimensional one, not ten- or more-dimensional as in some
modern physical models) is a special quantized medium arising as
the result of interaction of byuon's vacuum states (VSs). That is,
space is not given but arises. Therewith the appearing
three-dimensional space originated must have an insignificant
global anisotropy, as distinct from all basic isotropic models
with the same properties in various directions. The said
anisotropy denotes the existence of some chosen direction caused
by the existence, in nature, of a new fundamental vectorial
constant, the cosmological vector-potential $A_g$ entering into
the definition of the byuon. That new constant is associated with
the prediction of a novel anisotropic interaction of natural
objects between themselves and with the physical vacuum, a lowest
energy state of physical fields.

It should be noted that in the literature spaces with local rather
than global anisotropy are considered [41], for example, the
Finsler's space-time [42], but the local anisotropy is given
therein "by hand", as the saying goes. That is, an author himself
directively introduces it into his model instead of obtaining from
some general principle. For example, there are domain models of
the Universe.

Secondly, the physical sense of time notion is not yet revealed in
science in the present state of the art [43]. The general
philosophic concept of time as a form of matter existence, which
form expresses the order of change of objects and phenomena as a
sequence of events, does not indicate a common nature of those
events. As a rule, people tie their time to a particular periodic
process: rotation of the Earth around its axis, Earth's orbiting
around the Sun, oscillations of a quartz system, asf, without
becoming aware of inner, profound sense of time. Standard physical
time references, for example, quantum or, what is the same, atomic
clock with instrument error on the level of $10^{-11}$ per year
and moderate resolution on the order of $10^{-13}$ seconds, give
us no possibility of approaching the knowledge of time essence.
The byuon theory reveals physical essence of time as a discrete
sequence of changes in the byuon's "length", its quantum number. A
possibility therewith arises, to synchronize clocks at great
distances comparable with dimensions of our Metagalaxy, due to the
quantum process of physical space formation from the byuon's
vacuum states (VSs). That possibility substantially distinguishes
the theory of byuon's from A. Einstein's special theory of
relativity (STR), in which clocks can be synchronized only when a
signal has passed between them with speed of light $c_o$. It
should be noted at once that in the byuon theory, material objects
cannot move with a speed faster than the light speed (that is
similar to the STR's postulate on finite propagation velocity of
interactions), but synchronization of clocks occurs by a quantum
way without introducing the concept of speed. That is, some object
originated in the course of interaction between byuon vacuum
states and forming the physical space, is at a time in  two
spatial regions being very distant from each other in the
three-dimensional space arising.

Third, an essential distinction of the byuon theory from modern models in the classical and quantum field theories [44] is that the potentials of physical fields (gravitational, electromagnetic, asf.) become, in the theory of byuons, exactly fixable, measurable values. Recall to the reader that ordinary methods of measurement are capable to measure solely a difference of potentials. Therefore, in the existing field theory, potentials are defined only with a precision of an arbitrary constant or the rate of change of the potentials in space or time (gauge models). But in the theory of byuons, field potentials become single-valued since there are formed, on the set of byuon VSs, field charge numbers which generate the fields themselves, as, for example, the electric charge of an electron generates an electric field. The physical sense of field as a special form of matter, loses its basic meaning because all the observable events can be described on the basis of the byuon theory without introducing the concept of force, and hence of field.

An important methodic distinction between the byuon model and all those existent in the theoretical physics of today, is that the latter use images with properties of real objects, - for example, strings in the physics of elementary particles [45], superstrings, membranes when creating a unified field theory [46], asf. The byuons but are unobservable objects having no analogues in the nature though all the natural objects appear in the result of interaction of byuon VSs.

The proposed pattern of formation of the observed space $R_3$ on
the basis of dynamics of the finite set of byuons animates, fills
with a sense, and supplements the physical results on properties
of elementary particles, described in [34-35]. For example, if
some elementary object appearing in byuon interaction has, with
the probability near $1$, the vacuum state $I^+$ of a byuon
completing formation of its quantum numbers (the greatest period
of byuon interaction of the order of $kN$), such an elementary
object will be stable as well as its properties will, since quite
a definite amount of information will be locked up by VS $I^+$.
This relates, for example, to the electron.

Thus, as opposed to gauge models in which the level of symmetry
constantly grows for more complete, all-embracing, and unified
description of the surrounding world [23-28], and to obtain
massive particles it is necessary to use "by hands" the Higgs
mechanism (spontan violation of symmetry), in the present model
there exists first a one-dimensional world (its direction, i.e.
that of $\mathbf{A}_g$, is determined by the byuon with the
maximum $x(i)$), then its symmetrization takes place, and the
space $R_3$, with the world of elementary particles originates. At
that   some insignificant ($\sim 1/k$ ($\sim 10^{-15}$)) asymmetry
of "empty" $R_3$ remain as well as that of the $10^{-5}$ order
inside the elementary particles.

It should be also noted that to calculate the fundamental
constants $h, \, e_0, \, c$; constants of known interactions,
masses of main baryons, leptons, and mesons according to formulae
of the byuon theory, only three numbers $\tilde x_0, \, \tau_0, \,
|\mathbf{A}_ g|$ should be given since the characteristic
dimensions $\tilde x_0 \approx 10^{-17}$ ~cm, $c t^* \approx
10^{-13}$ ~cm and $10^{28}$ ~cm are found from the minimum PE of
byuons and from the information theorem.

Notice that in [20-22, 24,25, 31-38], results of some fundamental experiments in support of the basic theoretical statements are shown.

\section*{Appendix 3. Force-free physics. A qualitative pattern of a common approach to unifying all interactions. A novel principle of relativity.}

The byuon theory is presently in a stage of development but it is
seen from already put milestones of the novel physical concept
that the scope of this theory is much wider as compared to the
unified field theory because the former gives physical
interpretation of the observable space, reveals the nature of
notions of time, mass, charge, and so forth. But what is the
matter with a classical problem of physicists, - construction of a
unified approach to all interactions? The theory of byuon solves
this problem in an unusual way, namely, by way of full abandoning
the concept of force itself, that is, by  way of construction of
force-free physics. What usually sees an observer when
investigating one or another interaction between objects? He sees
the motion of objects which are either approaching each other, or
moving farther apart, or revolving one around another at a common
distance. For example, at a gravitational interaction of objects
we see that they approach each other, but at an electrical
interaction of two electron, for instance, they move away from
each other. During rotary movements of natural objects, - for
example, the Larmor rotation of electron in magnetic field,
revolution of planets around the Sun, own rotations of objects
asf., - the distances between the objects do not change. In terms
of forces, the phenomenon in consideration may be described by the
Newton's law of gravitation, Coulomb's law, Lorentz's force, asf.
But there is another approach at which all existent interactions
of objects in nature (four known and the fifth novel ones) are
explained by  way of variation of fundamental scales $x_0$ and $c
t^*$. Let us show it, but first recall how the concept of force is
introduced in  physics. At school we towed a weight by spring
along an inclined plane and measured the magnitude of frictional
force from tension of the spring. All the ordinary "everyday"
forces can be always reduced to the four known interaction
mentioned above. For example, the forces of friction, gas or
liquid pressure, forces in solid-state physics, - all are the
electromagnetic forces.

In the field theory, the concepts of "force" and "field" are
closely related. Let us give an extract from Ref. [47] on this
subject:

"The interaction between particles is explicable by the notion of
force field. Rather than say about the action of one particle onto
another, we can say that the particle creates a field around
itself, and any other particle being in that field, is subjected
to the action of some force. In the classical mechanics, the field
is only a means of describing the physical phenomenon of
interaction between particles. But in the theory of relativity the
situation substantially reverses because of a finite propagation
velocity of interactions. Force acting upon a particle at an
instant of time, do not determined by their arrangement at that
same instant. A change in position of one particle reflects on
other particles only after some time. That is, the field itself
becomes physically real. We may not say that the particles being
some distance apart, directly interact with each other. The
interaction can occur at each time point only between neighbouring
points in space (short-range interaction). Therefore we should
speak about interaction of a particle with a field and subsequent
interaction of the field with another particle."

At our approach, the concept of "field" is not necessary since the
dynamics of observed objects can be explained by the dynamics of a
finite set of VSs of a single object, the byuon. Let us show this
qualitatively for the existent types of interactions.

\textbf{Weak interactions}. In papers cited we have touched on
this subject when considered the  leptons and indicated that the
muon and tau-lepton decay under the action of weak interaction. In
the framework of the byuon theory, this interaction can be
explained in the following way. According to the basic hypothesis
of said theory, the entire world has come into being in
correspondence with the minimization of byuon VSs potential
energy. Only those particles can be stable which are corresponded,
when originating (i.e. forming their charge numbers), by minimum
periods of interaction of byuon VSs in the process of minimization
of potential energy, that is, the values $x_0$ and $c t^*$. For
muon and tau-lepton, the process of their origin is extended over
$2 x_0$ and $8 x_0$ in the observable space, therefore the nature
tends, in accordance with minimization principle, to reduce all to
the minimum, i.e. to the periods $x_0$ and $c t^*$. Constants of
this interaction are exactly calculated in the theory of byuon.
The vectorial constant $\mathbf{C}_v$ characterizing local
anisotropic properties of weak interaction, can be written as
follows:

\begin{equation}
\textbf{C}_v = e_0 \textbf{A}_g 2 x_0^3
\end{equation}

Firstly, draw the reader's attention to the fact that no
superfluous parameters are introduced to calculate $\mathbf{C}_v$
and other properties of the surrounding world as well. All
quantities in Eq. (A3) can be found from minimum potential energy
when specifying quantum of space, quantum of time, and modulus of
$\mathbf{A}_g$. Substituting the above given values of parameters
into Eq. (A3), we obtain $C_v = 1.35 \times 10^{-49} \, erg \cdot
cm^3$. This value is very close to experimental results.

Explain the physical sense of Eq. (29). The quantity $e_0 \times
\mathbf{A}_g$corresponds to the interaction of the electric charge
of a particle, in the process of its origin, with the vector
$\mathbf{A}_g$ during extension of the volume in which the
interaction takes place, up to $2 x_0^3$. That is, the object 4b
corresponding, for instance, to the muon, in the process of muon
generation will be in the volume $2 x_0^3$ with probability 1 (In
the common physics such an object is corresponded by a pair "muon
neutrino -- muon antineutrino"). Earlier one of the authors
(Yu.B.) in his books has also interpreted the weak interaction as
the loss of independence of spatial from time coordinates in some
generalized space, for example, in four-dimensional Minkovskian
space [20-21].

\textbf{Strong interactions}. At present, the majority of
physicists consider the theory of strong interaction as already
developed. This discipline is called Quantum Chromodynamics (QCD).
Its basic statements substantially supplement earlier models
advanced by Yukawa in 1934 and independently by Gell-Mann and
Zweig in 1964. The former is based on the existent particles
$\pi$-mesons being some interchange particles between protons and
neutrons in the nucleus (these particles fasten the protons and
neutrons into a single whole). The latter is a quark model.
According to it, hypothetical quarks should have unusual
properties: fractional charges (baryon charge $1/3$, electric
$+2/3 e_0$ and $-1/3 e_0$). To experimentally detect the quarks
assumed, immense financial means and intensive efforts of
physicists were spent, but all was in vain despite the energy of
particles colliding in accelerators reached enormous magnitudes of
thousands billions electron-volts. Then the theorists proposed a
string mechanism of interaction between quarks, so called QCD,
that implies that the more are quarks stretched, the stronger is
the interaction between them. That is, quarks in principle cannot
leave the nucleus. This problem was denoted as "confinement". In
experiments, "strings" longer than $10^{-13}$ cm, are not
detected.

Now about what the byuon theory speaks on the subject. The
existence of strong interactions is considered there as
interaction between the pairs of byuon VSs $I^+$ and $I^-$. Such
interpretation has an analogue in the classical electrodynamics
and quantum mechanics where the interaction of two electric
dipoles is considered. Their potential of interaction (Van der
Waals forces) depends on distance as does the Yukawa's potential,
that is, it can ensure basic properties of nuclear forces.
Therewith if charges of such dipoles will be one order of
magnitude larger than the electric charge, the constant of these
interactions will be about 1, as in the experiments [20-22]. In
the theory of byuon, the interaction of pairs of byuon VSs $I^+$
and $I^-$ participant in formation of the baryon charge of proton
and neutron, leads to a decrease in the period $c t^*$ of
interaction of byuon VSs (around $10^{-13}$ cm), which is
naturally manifested in attraction of nucleons (protons and
neutrons). The fact that the hypothetic objects, the quarks, do
not stretched into strings longer than $10^{-13}$ cm, now can be
qualitatively explained. That is clear since the next (after
$10^{-13}$ cm) period of interaction of byuons VSs is equal to
$10^{28}$ cm. That is, there exist no stable quantum structures
with dimensions between $10^{-13}$ cm and $10^{28}$ cm.

\textbf{Electromagnetic and gravitational interactions. } It
should be noted that the most close to our method of attack the
problem of searching unified pattern of interactions are works by
H.Weyl [48-49] and P.Dirac [50]. Historically, Weyl's geometry was
proposed to explain by space curvature not only long-range
gravitational field, as in the Einstein's theory, but another
long-range field, the electromagnetic field, too.

The gravitational field is well explained by Einstein's theory
that represents it in terms of space curvature. This has led to a
thought that the electromagnetic field also may be explained by
some property of space instead of an idea of its "immersion" into
the given space. It was thus necessary to build up some more
general space, as compared with Riemannean space forming the
foundation of Einstein's theory, which space would describe the
existent gravitational and electromagnetic forces and lead to
unifying long-range forces.

The space curvature required of Einstein's theory, can be
expressed in terms of parallel shift of a vector when moving it
along a closed contour, which leads to a difference between the
initial and final directions of the vector. Weyl's generalization
consisted in an assuming that the finite vector has not only
another direction but another length, too (and that is the most
important). In Weyl's geometry, there is no absolute means of
comparing length elements at two different points if they are not
infinitely close to each other. The comparison can be made only in
relation to a line segment connecting both points, and different
ways will give different results for the ratio of two length
elements. To have a mathematical theory of lengths, we must
arbitrarily specify standards of length at each point and then
correlate any length at any point with a local standard for this
point. Then we obtain some magnitude of vector length at any
point, but this magnitude is changed when is changed the local
standard of length.

Hence the Weyl's geometry ensures just what is necessary for
describing the gravitational and electromagnetic fields in
geometric terms, that is, by changes in periods of byuon VSs
interactions (fundamental scales). In the works [20,21,23] is
shown that by way of changing fundamental scales, we can obtain
Maxwell's equations describing properties of the electromagnetic
field. Let us show the gist of the matter on a qualitative level.

When assumed that there is composed a mathematical algorithm for transition of events from the one-dimensional space $R_1$ to three-dimensional $R_3$, the electromagnetic and gravitational interactions of objects can be achieved by changing periods of interaction of byuon VSs $k$, $N$ and, respectively, by changing scales $x_0$ and $c t^*$. The electrostatic field is caused by a change in $|\mathbf{A}_{\Sigma}|$ due to some scalar $A^0$, the magnetic field is due to vectorial potential $\mathbf{A}$ but with complete return (without deviations) of the vector $\mathbf{A}$  to the initial point of space $R_3$ after circulation along some closed contour $\ell$ (in that case  $\cos _{II^+ II^+} = \cos _{II^- II^-} = 0$). It should be noted once more that the electromagnetic constant of interaction (constant of thin structure) $e^2/\hbar c$ will be the same in all reference systems.

Let us illustrate the above said by the example of static electric and gravitational fields. For an electron in electrostatic field with potential $A^0$ we can write the following expression

$$
m_e c^2 + e A^0 = m'_e c^2 = 2 m_{\nu_e} N'.
$$

Here we assume the electric field to be absent as such, but there are only changes in periods of byuon VSs interaction corresponding to the effect of presence of electric field at formation of this electron. That is, in fact only the own mass of electron is changed.

From the above relationship is seen that if $e A^0 > 0$ then $N' >
N$ (starting value of the period before appearance of $A^0$), and
$N' < N$ if $e A^0 < 0$. That is, the interaction of like charges
leads to repulsion of them (increase in the period $N$), and
unlike charges are attracted together (decrease in $N$).

For the electron in a static gravitational field with potential
$\varphi$ an analogous relationship can be written: $m_e c^2 + m
\varphi = m_e'' c^2 = 2 m_{\nu_e} N'' $.

(Here we also assume that there is no gravitational field as such
but only the change in periods of interaction of byuons VSs
correspondent with the effect of presence of gravitational field
at formation of the electron). In this case we deal only with
attraction since $\varphi < 0$, and hence $N'' < N$.

\textbf{New interaction}. We name this interaction as byuon
interaction. That is principally distinct from the existent
interactions since is connected with global anisotropy of space
caused by the vector $\mathbf{A}_g$ appearing in the definition of
byuon $[A_g x(i)]$. The distinction is in that the functions $\cos
_{II^+ II^+}$ $\cos _{I^- II^-}$, $\cos _{II^- II^-}$ asf, do not
precisely equal to zero, and there is going the process of
extension of byuon in vacuum states $II^+$ and $II^-$, that is,
space is coming into being. In terms of a simple geometric model,
for any limited volume of $R_3$ always a vector of byuon
$\mathbf{A}_{\Sigma}$ can be found that enters the volume at some
point of the surface embracing the volume, at the minimum angle of
$\sim 1/10^{15}$, and another vector of byuon always comes out of
the volume nearly at the same angle, creating global anisotropy of
the Universe. Therewith the change of $\mathbf{A}_{\Sigma}$  in
$\mathbf{A}_g$ direction due to the vector potential of a magnetic
system also changes the period of byuons VSs interaction $k$, i.e.
the scale $x_0$, and leads to rejection of an object out of the
region of decreased $|\mathbf{A}_{\Sigma}|$.

The new interaction, contrary to others, is of sharply anisotropic character and directed along generatrixes of above mentioned cone to the side of vector $\mathbf{A}_g$, but that interaction is zeroth in the region of precise antiparallelism of the potential $\mathbf{A}$ being generated, to the direction of $\mathbf{A}_g$, since the derivative $\partial \Delta A / \partial x = 0$ exactly along the vector $\mathbf{A}_g$.

\textbf{Unified pattern of all existent interactions} consists in
the fact that they are associated with one or another character of
changes in fundamental scales of physical space originated in the
process of byuon VSs interactions. These scale changes are
corresponded by one or another change of $\mathbf{A}_{\Sigma}$
that is taken in the traditional physics for the action on a
particle by one or other field. Thus in terms of traditional
physics, the byuon theory is a theory of "addition" of potentials
of all the existent fields with some fundamental potential
$\mathbf{A}_{\Sigma}$ that cannot be more than $|\mathbf{A}_g|$ in
magnitude. This is just the solution of the problem of unified
field theory. When speaking about the new interaction, we have
separated in it only anisotropic properties, but the quantity
$\mathbf{A}_{\Sigma}$ contains potentials of all known fields,
too. Therefore the theory of single force is sure to be created in
the nearest future on the basis of the new interaction, and from
that theory the expressions for all the known forces in nature
will be derived in one or another asymptotic approximation.

\textbf{Principle of relativity. }

 The phenomenon in
consideration, i.e. the new interaction, is nonlinear and
nonlocal. When getting to the root of its nature as well as nature
of other interactions associated with changes of fundamental
scales, we come to a force-free description of objects in the
world as well as to necessity of improvement of relativity
principle.

According to the relativity principle of Galilei,  all laws of nature are the same in all inertial systems of coordinates or, if more exactly, the expressions for these laws are invariant relative to transformation of coordinates and time when going from one inertial coordinate system to another.

The full invariance of all inertial reference systems points to
the absence of any "absolute" system of coordinates. This is the
essence of relativity of phenomena.

In the mechanics of Galilei the interactions propagate instantly, and time is absolute, that is, considered as independent of reference system.

Leaning upon the Michelson's experiment (1881) demonstrated light
speed independence of direction of its propagation, A.Einstein
developed the principle of relativity having supplemented it with
a postulate of finiteness of velocity of propagating interactions.
According to the relativity principle, the value of this velocity
is the same for all inertial reference systems and equal to the
light speed $c_0$. The limitation on propagation velocity of
interactions led to the fact that time ceased to be absolute and
became dependent upon the reference system, and the laws of
transition from one inertial coordinate system to another acquired
a non-Galileian form and were embodied in Lorentz's
transformations.

What a new is introduced in relativity of events in $R_3$ by the new approach to formation of $R_3$ from VSs of discrete object, the byuon, in accordance with the Hypothesis 1. To answer the question, it should be once more reminded of what the coordinate system is at all in such an approach, and after that determine what we will call the inertial system of coordinates.

Since we have put a question of how the observed physical space is
built, we already should not "arithmetize emptiness" as that is
made in the classic physics, quantum mechanics, and quantum field
theory. We can only find coordinates of some observed physical
objects relative to other objects.

For this reason alone the notion of coordinate system loses its
uniqueness. Further, if we do determine arrangement of objects
under observation relative each other with certain accuracy in the
context of large scaled world, we can approach the notion of an
inertial system of coordinates, too. By inertial systems of
coordinates such systems are meant in which the periods (k, N, P)
of byuon interactions (or fundamental scales $x_0, \, ct^*, \,
\tilde x_0 k N P$) as well as functions of $\cos _{II^+ II^+}$,
$\cos _{II^- II^-}$ type are the same. Of course, this statement
is fulfilled always with some accuracy. Therein does lie basic
idea of extension of the concept of relativity. In this connection
it should be noted that all forces in nature as well as the new
interaction will be zero in inertial coordinate systems (the
earlier definition of relativity principle) as they all can be
explained in terms of change of fundamental scales that do not
altered in such systems. Therefore in the experiments on
investigating the new force [20-22,32-35], the weight was always
suspended on a cantilever, if only on the order of 1cm in length,
since otherwise the new force is zeroth when the body and sensor
are in regions of the same scale $x_0$.

The force-free description of motion of objects gives an in-depth
understanding of nature, but the stereotype of our mentality and a
sufficient accuracy of description of motion of objects in the
everyday life (by using the second Newton's law) lead to the
notion of the new force though physically that is only changes in
fundamental scales.

\begin{figure}[h]
   \includegraphics{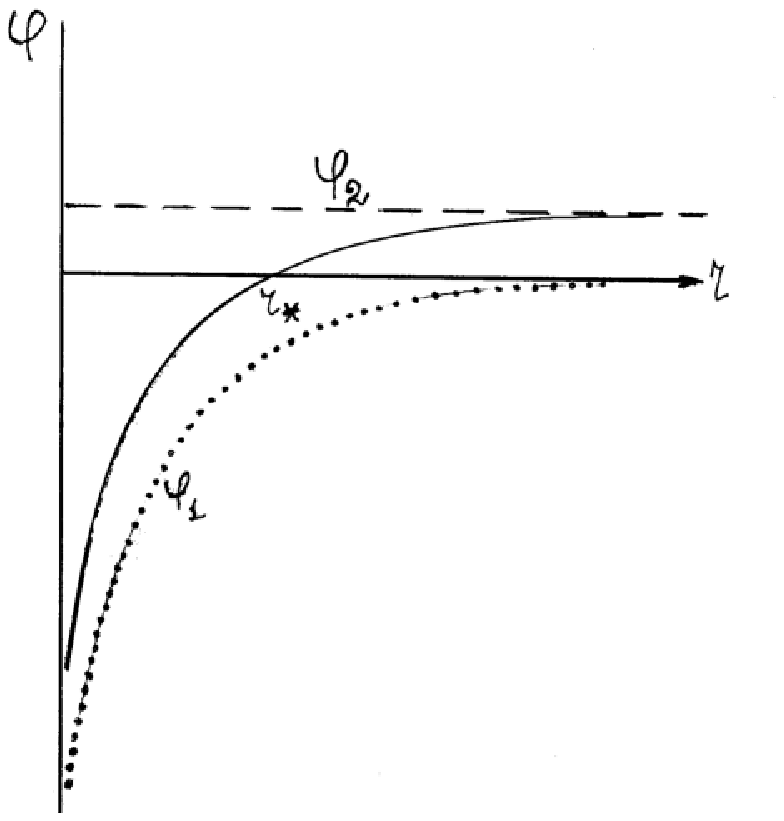}
   \caption{Scheme of two body interactions in our Universe.}
\end{figure}

\begin{figure}[h]
   \includegraphics{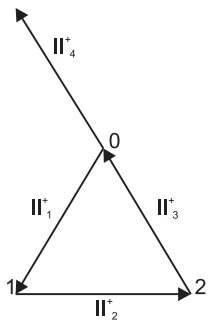}
   \caption{Interaction of byuons in vacuum states  $II_1^+$ and $II_4^+$  (the
smallest loop).}
\end{figure}

\begin{figure}[h]
   \includegraphics{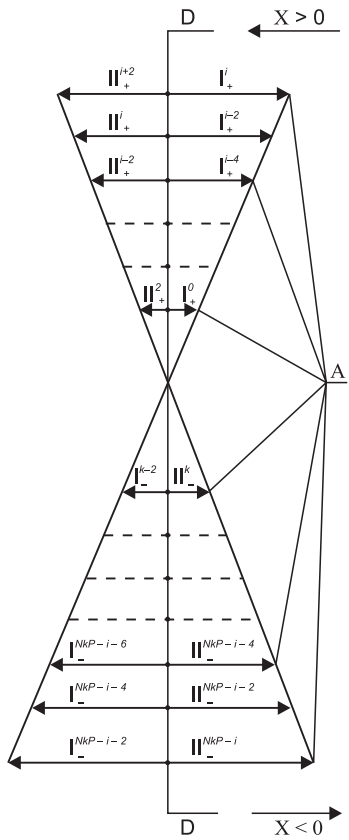}
   \caption{Completion of vacuum states $II^+$ and $II^-$ by vacuum states $I^+$ and $I^-$, respectively, at the maximum potential energy of interaction.
}
\end{figure}

\begin{figure}[h]
   \includegraphics{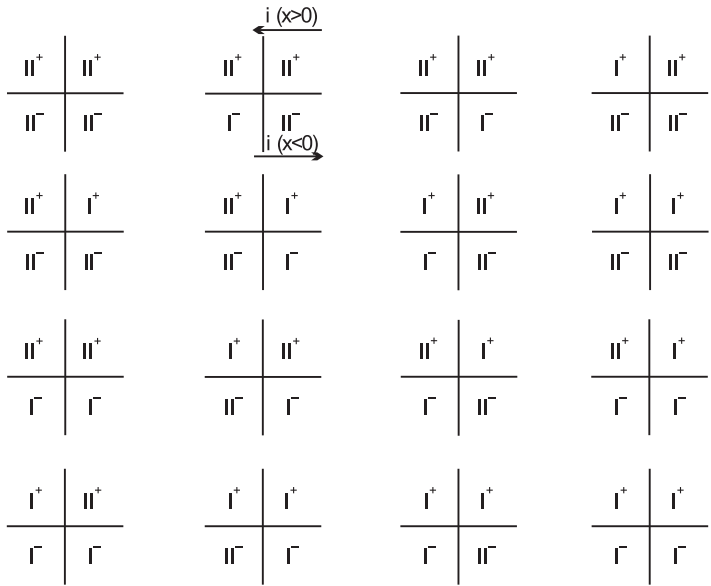}
   \caption{The possible variants of four-contact interaction of byuons. The interaction being realized in nature is $II^+ - I^+ - II^- - I^-$.
}
\end{figure}

\begin{figure}[h]
   \includegraphics{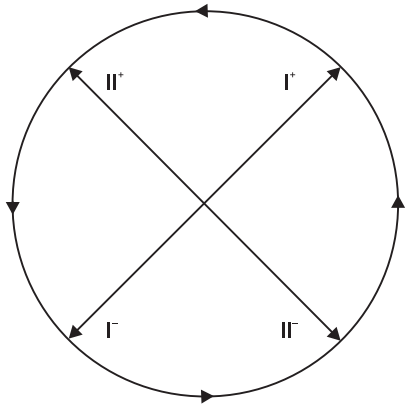}
   \caption{The rule of circular arrow for determining the distance between byuons.
}
\end{figure}

\begin{figure}[h]
   \includegraphics{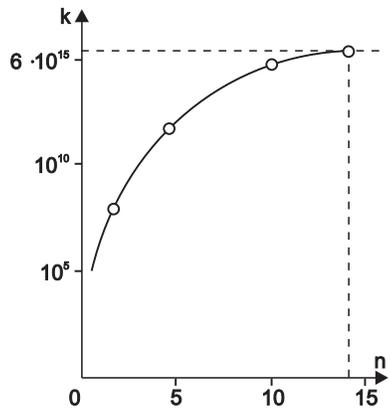}
   \caption{k as a function of the number n of  the terms of the series.
}
\end{figure}

\begin{figure}[h]
   \includegraphics{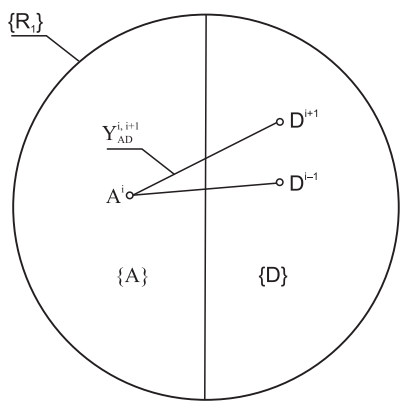}
   \caption{Representation of the set $\{R_1\}$ as a union of sets of the points $\{A\}$ and $\{D\}$.
}
\end{figure}
\end{document}